\documentclass[preprint,preprintnumbers,amsmath,amssymb]{revtex4}

\usepackage{graphicx}% Include figure files
\usepackage{dcolumn}% Align table columns on decimal point
\usepackage{bm}% bold math
\begin{document}

%\preprint{APS/123-QED}

\title{On a separation criterion for symmetric elliptic bluff body flows}
\author{Kaushik Srinivasan}
 \email{kaushik@jncasr.ac.in}
\affiliation{Engineering Mechanics Unit,
Jawaharlal Nehru Centre for Advanced Scientific Research, Jakkur,
Bangalore 560064, India \\
}
%------------------------------------------------------------------------------

\date{\today}

\begin{abstract}
A new analytical criterion that captures the onset of separation of flow 
past elliptic cylinders is derived by considering the variation of the wall
normal velocity in Reynolds number parameter space. It is shown that this 
criterion can be used to calculate the separation Reynolds number 
($Re_{s}$) for the classical problem of flow past a circular cylinder,
a contentious and unresolved issue till date. The two dimensional 
Navier-Stokes equations are solved computationally and an exact value of  
$Re_{s}$ is obtained by applying the aforementioned criterion.
\end{abstract}

\maketitle	
%------------------------------------------------------------------------------

A great deal of research in the past century has been focused on bluff body 
wakes and considerable progress has been made towards understanding them.
A body is considered ``bluff" if the spatial extent of the body along the 
flow direction is of comparable or lesser order to that normal to it. 
The  bluff bodies that are of concern here are a family of symmetric 
elliptical cylinders, with their minor axes aligned with the flow direction. 
The flow past such a bluff body is steady for very low Reynolds numbers 
($Re=Ua/\nu$ where $a$ is the length of the body normal to the flow 
and $\nu$ is the coefficient of viscosity), which in case of a circular 
cylinder happens for $Re<47$. In two dimensions, as $Re$ is increased, 
a few well defined features of the flow are observed. Up to a particular 
value of $Re$ (say $Re_{s}$), the streamlines, while being asymmetric 
about the axis normal to that of the flow, are attached to the body 
\cite{trit}. But for $Re>Re_{s}$, the flow separates and two well defined 
separation ``bubbles" or eddies are observed \cite{trit}. These ``bubbles" 
are regions of closed streamlines where the flow direction just next to 
the cylinder is in the opposite direction to the mean flow ($i.e.$ a 
region of backflow). The characteristic  features of these bubbles have 
been documented comprehensively both in computational and experimental 
studies, mostly for the case of flow past a circular cylinder. 
Investigations of flow past elliptic cylinders are few \cite{bala} and 
none of them seem to focus on aspects of flow separation. Computational results 
\cite{dennis,fornberg} indicate that for the case of a circular cylinder, 
$4<Re_{s}<7$, but even recent investigations, for example \cite{wu2},  
have been unable to capture the bubble for $Re<6$ inspite of having a 
much higher numerical resolution in the wake region than \cite{dennis,fornberg}.
Experimental studies \cite{nisi,taneda,bouard} suffer from similar ambiguities 
and an effort to summarize and compile various such experiments  was undertaken 
by \cite{prup} who concluded that $Re_{s}\sim 5$. The criterion derived here,
as we shall see later, can be used to obtain the exact value of $Re_{s}$ and 
resolve this ambiguity.

The complexity of the problem means that only a few comprehensive theories 
have been put forth in the past and their scope and success have been limited. 
One theory of note by \cite{smith1,smith2,smith3} uses the triple-deck model 
to show that the length of the bubbles (in the direction of the flow) increases 
linearly with Reynolds number. This theoretical result has been confirmed by 
many experiments and numerical simulations \cite{fornberg2}. 
However, this theory is unable to predict the Reynolds number at which the 
bubbles appear, $Re_{s}$. Further, it does not attempt to explicitly understand 
and characterize the flow when bubbles form. Neither does it attempt to 
generalize the same for other bluff bodies like  symmetric elliptic cylinders 
in which the bubble formation is of a similar nature.

The aim of this Rapid Communication is an attempt, a possible first 
step, in this direction. Here we derive a simplistic analytical criterion 
that captures 
the Reynolds number at which the bubbles start forming, $i.e.$ $Re_{s}$.
This criterion is general and valid for a whole class of elliptic cylinders.
It seems similar to the Prandtl shear stress criterion which states that,
at the point where the streamline separates the wall shear stress, 
			$\tau_{wall}=0$.
The distinction between the Prandtl criterion and the present one though is
quite fundamental and should pose no source of confusion. The criterion 
derived here is valid {\sl only} at $Re_{s}$, $i.e.$ it captures
the Reynolds number at which the bubble starts forming. Whereas once 
the bubble forms, the upper and lower streamlines ending at the wall 
(the ``separatrices" since they separate the ``outer" flow region
from the flow ``inside" the bubble) always satisfy the Prandtl shear 
stress criterion at the point that they separate, irrespective of the Reynolds
number (of course, until the onset of unsteadiness).
The Prandtl criterion, it must be noted, has relevance only {\sl after} 
the bubble is formed and the point that the separatrices separate from the
cylinder surface has $\tau_{wall}=0$, but this has absolutely no bearing on
$Re_{s}$.

We study uniform flow past a 2D elliptic cylinder such that the flow direction
is along the minor axis, as shown in  Fig. 1. A body-fitting orthogonal 
coordinate system given by $\zeta-\eta$ (specifically, an elliptic 
cylindrical coordinate system) coincides with the $x-y$ axis at the 
point P, the intersection of the symmetry plane with the cylinder. One expects
that, as the $Re$ is increased from zero through $Re_{s}$, it is at P that
the bubbles start forming. Examining the flow in the vicinity of P in $Re$ 
parameter space would give us the required criterion. This is done by considering
the flow around P for two Reynolds numbers $Re_{1}$ and $Re_{2}$ such that 
\begin{equation}
0<Re_{1}<Re_{s}<Re_{2}<Re_{uns}
\end{equation} 

\begin{figure}
\includegraphics[width=0.32\textwidth]{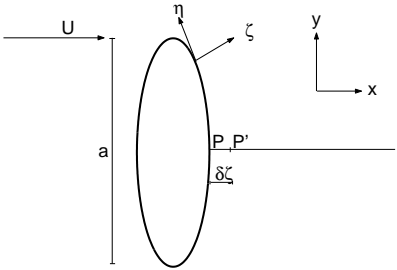}
\caption{Schematic of the flow geometry}
\label{schema}
\end{figure}
%------------------------------------------------------------------------------
Here $Re_{uns}$ is the Reynolds number at which the steady wake of the 
cylinder becomes unstable and vortex shedding sets in, $e.g.$ $Re_{uns}=47$ 
for a circular cylinder. Now for $Re=Re_{1}$, the bubble has not started 
forming and there is no region of backflow; whereas for $Re=Re_{2}$, the 
bubble has formed and is of finite size and has a region of backflow.
Then one can find a distance $\delta \zeta$, which is less than the bubble 
length, so that,
\begin{eqnarray}
u(\zeta_{P}+\delta \zeta,\eta_{P},Re_{2})<0 \\
u(\zeta_{P}+\delta \zeta,\eta_{P},Re_{1})>0
\end{eqnarray} 
Here $P\equiv(\zeta_{P},\eta_{P})$, $P'\equiv(\zeta_{P}+\delta\zeta,\eta_{P})$ 
and the flow velocity in $\zeta-\eta$ coordinates is written as 
$u\equiv u_{\zeta}$ and $v\equiv u_{\eta}$ along 
the respective coordinates.

Note that the no slip condition at the wall implies that u$|_{P}$=v$|_{P}$=0. 
Since $v=0$ along the cylinder, $\frac{\partial v}{d\eta}|_{P}=0$ (so are 
derivatives to higher orders along that direction). Using the continuity 
equation at P, 
$\frac{1}{h_{\eta}}\frac{\partial v}{\partial{\eta}}+\frac{1}{h_{\zeta}}
\frac{\partial u}{\partial{\zeta}}=0$ (here $h_{\eta}$ and $h_{\zeta}$ are 
the scale factors for the coordinate transformation (x,y) $\rightarrow$ 
($\zeta,\eta$) ) and the fact that $\frac{\partial v}{d\eta}|_{P}=0$, we get
\begin{equation}
\biggl.\frac{\partial u}{\partial{\zeta}}\biggr|_{P}=0
\end{equation} 
%------------------------------------------------------------------------------

Now we write the $u$ velocity at point $P'$ as a Taylor expansion about the 
point $P$, keeping terms till second order. Since $\delta\zeta$ is along the 
$\zeta$ direction, we get

\begin{equation}
u|_{P'}=u|_{P}+ \biggl.\frac{\partial u}{\partial{\zeta}}\biggr|_{P} \delta 
\zeta +\biggl.\frac{\partial^{2} u}{\partial{\zeta}^{2}}\biggr|_{P}
\frac{{\delta \zeta}^2}{2!}+O(\delta \zeta^3)
\end{equation} 
From no slip and Eq. (4) we get
\begin{equation}
u|_{P'}=\biggl.\frac{\partial^{2} u}{\partial{\zeta}^{2}}\biggr|_{P}
\frac{{\delta \zeta}^2}{2!}+O(\delta \zeta^3)
\end{equation} 
Clearly, the relations (2) and (3) imply that $u|_{P'}(Re)$=0 at some 
$Re_{1}<Re<Re_{2}$. 
By letting the limits $Re_{1}\rightarrow Re_{s}^{-}$ and $Re_{2}\rightarrow 
Re_{s}^{+}$ alongwith $\delta \zeta\rightarrow0$, we see that this Reynolds 
number is precisely $Re_{s}$, $i.e.$ $u|_{P'}(Re_{s})$=0. And from (6) we 
finally get,
\begin{equation}
\frac{\partial^{2} u}{\partial{\zeta}^{2}}(\zeta_{P},\eta_{P},Re_{s})=0
\end{equation} 
which is precisely the criterion required for separation. The above derivation
is almost trivial and the result is fairly intuitive. Further, it is seen that 
the condition (7) is necessary and sufficient. This is because 
$u$ near P is  always positive for $Re<Re_{s}$ and so is the second derivative.
Similarly for $Re_{uns}>Re>Re_{s}$ it is always negative, again implying a 
negative value of the second derivative. An equivalent pressure condition  
can also be derived by using (7) and considering the Navier-Stokes 
(NS from here on) equation at P. Setting all derivatives of  velocity 
components along the wall to be zero and using Eq. (4), we get,
\begin{equation}
\frac{1}{h_{\zeta}}\frac{\partial^{2} u}{\partial{\zeta}^{2}}
(\zeta_{P},\eta_{P},Re)=Re\frac{\partial p}{\partial{\zeta}}(\zeta_{P},
\eta_{P},Re)
\end{equation} 

for $Re=Re_{s}$, we have from (7), 
\begin{equation}
\frac{\partial p}{\partial{\zeta}}(\zeta_{P},\eta_{P},Re_{s})=0
\end{equation} 
Of the two equivalent conditions, (7) and (9), it is not clear which would 
be more useful in computational efforts in determining $Re_{s}$. 
This can be resolved by examining the behavior of $\frac{\partial^{2} u}
{\partial{\zeta}^{2}}(\zeta_{P},\eta_{P},Re)$ and $\frac{\partial p}
{\partial{\zeta}}(\zeta_{P},\eta_{P},Re)$ in $Re$ space around $Re_{s}$.

It is instructive to note that relations similar to (7) and (9) would
also be valid in three dimensions for flow past a family 
of symmetric ellipsoids. This family, to which the sphere also belongs,
has rotation symmetry about an axis parallel to the flow direction.
The derivation of the separation criteria for this family mirrors the 
one above. While flow past an ellipsoid has not been investigated in 
any detail, for a sphere, experiments imply that $Re_s$=25 
\cite{taneda2,johnson}. However, unlike the case of a flow past
a cylinder, there is excellent agreement between various experimental and
computational results \cite{johnson} about this value. In the light
of this fact, the criterion (7) has little application here. Our present 
study is therefore restricted to the study of two dimensional flows.

We consider flow past a circular cylinder, which is a special case of the 
family of ellipses and relatively easy to compute. In this case, $(\zeta,\eta)
\equiv (r,\theta)$ but we continue to use $(\zeta,\eta)$ for the sake of 
uniformity of notation. We  solve the 2D NS equations computationally in the 
streamfunction-vorticity formulation. This approach has an advantage 
over primitive variable formulations since an explicit operator 
splitting is not required.
The unsteady NS equations are solved in the range $2.5<Re<10$ by impulsively
starting the flow, unlike the authors \cite{fornberg} who solve the steady 
NS equation. Impulsively started flow past a cylinder has been studied 
extensively using a variety of numerical formulations - finite difference,
finite volume and vortex methods \cite{leonard}. The present work uses a 
finite difference formulation for its simplicity as done by \cite{dennis}.
It is formulated using an explicit time stepping scheme,
the time stepping being varied from a fourth Runge Kutta scheme
to an Euler scheme with identical results. The discretization of 
the non-linear terms was done using a 3rd order upwind scheme \cite{kuwahara}
while all other derivatives were effected using central differencing. 
The streamfunction-vorticity Poisson equation was solved using a stabilized
Biconjugate Gradient Method \cite{saad}. The grid used was a body fitting grid,
with grid clustering in the radial direction. Computations were performed on 
three different grids with the highest resolution being $300 \times 150$ 
and the largest ratio of the outer boundary to the cylinder diameter being 40.
Each computation was performed till steady state was reached. About a hundred
hours of computation time were required for all the cases on the most refined
grid, the results of which are presented below.

\begin{figure}	
\includegraphics[width=0.43\textwidth]{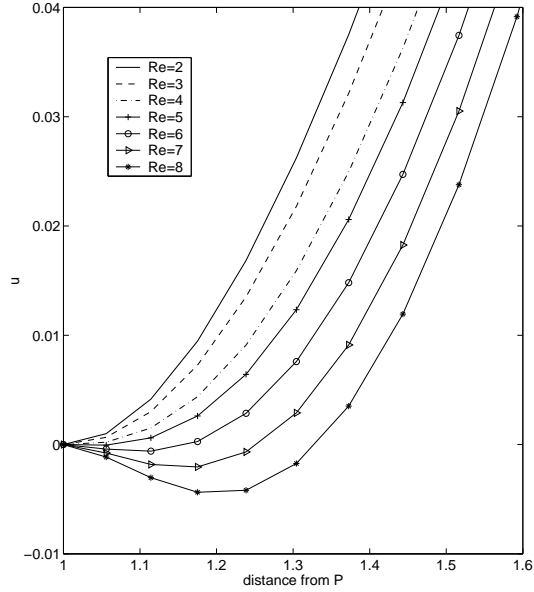}
\caption{Variation of the $u$ velocity along the symmetry line starting from P
at different Reynolds Numbers.}
\label{prim1}
\end{figure}
Once the flow is started impulsively from rest, a pair of separation bubbles 
can be seen for $Re>6$, below which it was hard to resolve the bubbles. The 
computations were continued up to a time t=10 (where $U=\frac{1}{2}D=1$, 
$U$ and $D$ being the free stream velocity and cylinder diameter respectively)
when steady state was approximately reached. The difficulty in resolving the 
bubbles close to separation is not a new problem and is common to computational
investigations of this nature\cite{wu2}. In order to do this, we first look 
at the $u$ velocity variation along the symmetry plane close to the cylinder.
This has been done and the results are plotted in Fig. 2 for a range of 
Reynolds numbers as indicated. Clearly $u<0$ near P indicates the presence 
of a bubble. As seen from Fig. 1, $4<Re_{s}<6$, which is in good agreement 
with existing computational results. 

\begin{figure}
\includegraphics[width=0.38\textwidth]{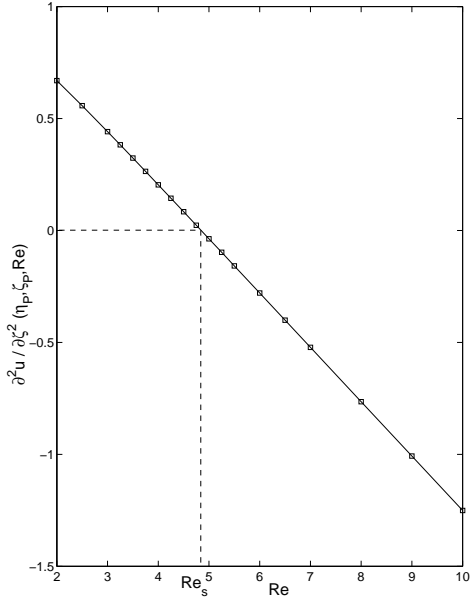}
\caption{Variation of $\frac{\partial^{2} u}{\partial{\zeta}^{2}}(\zeta_{P},
\eta_{P},Re)$ with Reynolds number at the point P.}
\label{veltf}
\end{figure}
In Fig. 3, $\frac{\partial^{2} u}{\partial{\zeta}^{2}}(\zeta_{P},\eta_{P}
,Re)$, which is obtained from the $u$ velocity data by central differencing,
is plotted against Reynolds number. As seen in Fig. 3 we obtain the rather 
surprising result that $\frac{\partial^{2} u}{\partial{\zeta}^{2}}
(\zeta_{P},\eta_{P},Re)$ varies {\sl linearly} with $Re$ in the neighborhood 
of $Re_{s}$ and so, by virtue of (8), $\frac{\partial p}{\partial{\zeta}}
(\zeta_{P},\eta_{P},Re)$ does not. Therefore, numerical simulations 
attempting to compute $Re_{s}$ can do so by calculating $\frac{\partial^{2} u}
{\partial{\zeta}^{2}}(\zeta_{P},\eta_{P},Re)$ 
for a few Reynolds numbers around $Re_{s}$ and then extrapolating the curve 
linearly to zero to obtain $Re_{s}$. $\frac{\partial p}{\partial{\zeta}}
(\zeta_{P},\eta_{P},Re)$ on the other hand is highly non-linear around 
$Re_{s}$ and cannot be used in similar manner, even though (9) itself remains
valid. So from Fig. 3 and applying (7) one gets $Re_{s}$=4.8.
The aforementioned approach underlines the advantage of using (7) in order to 
get $Re_{s}$ as opposed to attempts in resolving the bubble by solving the NS
equation for a large number of values of $Re$. One must realize that, though 
(7) is exact, the linearity of $\frac{\partial^{2} u}{\partial{\zeta}^{2}}
(\zeta_{P},\eta_{P},Re)$ around $Re_{s}$ is observed {\sl only} through 
computations and  that too for the special case of a circular cylinder.
But from the {\sl seemingly} general nature of the problem of separation 
for the family of ellipses, as seen in deriving (7), one might expect that 
the linearity property holds well across the family. But an {\sl ad hoc} 
argument of generality can hardly suffice and therefore the author is 
currently investigating this aspect of the problem, both analytically and 
numerically. 

Also, a note of caution regarding the value of $Re_{s}$
obtained here, which it must be remarked is still contingent on the 
computational approach used. And the present computation while reasonable,
can be improved with respect to the grid refinement and also
the outer domain size, which might have some bearing on the value 
of $Re_{s}$. In fact there was some dependence of the exact value 
of $Re_{s}$ on the outer domain size (about 7$\%$ between 30 and 40 times the 
cylinder diameter for the most refined grid). But the linearity property 
remained unchanged and the variation in the slope of the 
$\frac{\partial^{2} u}{\partial{\zeta}^{2}}(\zeta_{P},\eta_{P},Re)$ 
curve was also negligible.

{\bf Acknowledgments:} The author wishes to thank Dr.~Rama Govindarajan and 
Prof. Roddam Narasimha for excellent critical comments and discussions 
and the former specially for also suggesting this problem. Thanks 
to Pinaki B., for valuable comments and discussions and, Komala R. and 
Dr.~A.~Sameen for help with the manuscript.

%\bibliography{myref_prl}

\begin{thebibliography}{18}
\expandafter\ifx\csname natexlab\endcsname\relax\def\natexlab#1{#1}\fi
\expandafter\ifx\csname bibnamefont\endcsname\relax
  \def\bibnamefont#1{#1}\fi
\expandafter\ifx\csname bibfnamefont\endcsname\relax
  \def\bibfnamefont#1{#1}\fi
\expandafter\ifx\csname citenamefont\endcsname\relax
  \def\citenamefont#1{#1}\fi
\expandafter\ifx\csname url\endcsname\relax
  \def\url#1{\texttt{#1}}\fi
\expandafter\ifx\csname urlprefix\endcsname\relax\def\urlprefix{URL }\fi
\providecommand{\bibinfo}[2]{#2}
\providecommand{\eprint}[2][]{\url{#2}}

\bibitem[{\citenamefont{Tritton}(1988)}]{trit}
\bibinfo{author}{\bibfnamefont{D.~J.} \bibnamefont{Tritton}},
  \emph{\bibinfo{title}{Physical Fluid Mechanics}} (\bibinfo{publisher}{Oxford
  University Press}, \bibinfo{year}{1988}), \bibinfo{edition}{2nd} ed.

\bibitem[{\citenamefont{Mittal and Balachandar}(1996)}]{bala}
\bibinfo{author}{\bibfnamefont{R.}~\bibnamefont{Mittal}} \bibnamefont{and}
  \bibinfo{author}{\bibfnamefont{S.}~\bibnamefont{Balachandar}},
  \bibinfo{journal}{J. Comp. Phys.} \textbf{\bibinfo{volume}{124}},
  \bibinfo{pages}{351} (\bibinfo{year}{1996}).

\bibitem[{\citenamefont{Dennis and Chang}(1970)}]{dennis}
\bibinfo{author}{\bibfnamefont{S.~C.~R.} \bibnamefont{Dennis}}
  \bibnamefont{and} \bibinfo{author}{\bibfnamefont{G.~Z.} \bibnamefont{Chang}},
  \bibinfo{journal}{J. Fluid Mech.} \textbf{\bibinfo{volume}{42}},
  \bibinfo{pages}{471} (\bibinfo{year}{1970}).

\bibitem[{\citenamefont{Fornberg}(1980)}]{fornberg}
\bibinfo{author}{\bibfnamefont{B.}~\bibnamefont{Fornberg}},
  \bibinfo{journal}{J. Fluid Mech.} \textbf{\bibinfo{volume}{98}},
  \bibinfo{pages}{819} (\bibinfo{year}{1980}).

\bibitem[{\citenamefont{Wu et~al.}(2004)\citenamefont{Wu, Yen, Wen, Chen, and
  Wang}}]{wu2}
\bibinfo{author}{\bibfnamefont{M.~H.} \bibnamefont{Wu}},
  \bibinfo{author}{\bibfnamefont{R.~H.} \bibnamefont{Yen}},
  \bibinfo{author}{\bibfnamefont{C.~Y.} \bibnamefont{Wen}},
  \bibinfo{author}{\bibfnamefont{M.~C.} \bibnamefont{Chen}}, \bibnamefont{and}
  \bibinfo{author}{\bibfnamefont{A.~B.} \bibnamefont{Wang}},
  \bibinfo{journal}{J. Fluid Mech.} \textbf{\bibinfo{volume}{515}},
  \bibinfo{pages}{233} (\bibinfo{year}{2004}).

\bibitem[{\citenamefont{Nisi and Porter}(1923)}]{nisi}
\bibinfo{author}{\bibfnamefont{H.}~\bibnamefont{Nisi}} \bibnamefont{and}
  \bibinfo{author}{\bibfnamefont{A.~W.} \bibnamefont{Porter}},
  \bibinfo{journal}{Philosophical Magazine} \textbf{\bibinfo{volume}{16}},
  \bibinfo{pages}{153} (\bibinfo{year}{1923}).

\bibitem[{\citenamefont{Taneda}(1956{\natexlab{a}})}]{taneda}
\bibinfo{author}{\bibfnamefont{S.}~\bibnamefont{Taneda}},
  \bibinfo{journal}{Journal of Physical Society of Japan}
  \textbf{\bibinfo{volume}{11}}, \bibinfo{pages}{302}
  (\bibinfo{year}{1956}{\natexlab{a}}).

\bibitem[{\citenamefont{Coutanceau and Bouard}(1977)}]{bouard}
\bibinfo{author}{\bibfnamefont{M.}~\bibnamefont{Coutanceau}} \bibnamefont{and}
  \bibinfo{author}{\bibfnamefont{R.}~\bibnamefont{Bouard}},
  \bibinfo{journal}{J. Fluid Mech.} \textbf{\bibinfo{volume}{79}},
  \bibinfo{pages}{231} (\bibinfo{year}{1977}).

\bibitem[{\citenamefont{Pruppacher et~al.}(1980)\citenamefont{Pruppacher,
  LeClair, and Hamielec}}]{prup}
\bibinfo{author}{\bibfnamefont{H.~R.} \bibnamefont{Pruppacher}},
  \bibinfo{author}{\bibfnamefont{B.~P.} \bibnamefont{LeClair}},
  \bibnamefont{and} \bibinfo{author}{\bibfnamefont{A.~E.}
  \bibnamefont{Hamielec}}, \bibinfo{journal}{J. Fluid Mech.}
  \textbf{\bibinfo{volume}{44}}, \bibinfo{pages}{781} (\bibinfo{year}{1980}).

\bibitem[{\citenamefont{Smith}(1979)}]{smith1}
\bibinfo{author}{\bibfnamefont{F.~T.} \bibnamefont{Smith}},
  \bibinfo{journal}{J. Fluid Mech.} \textbf{\bibinfo{volume}{92}},
  \bibinfo{pages}{171} (\bibinfo{year}{1979}).

\bibitem[{\citenamefont{Smith}(1981)}]{smith2}
\bibinfo{author}{\bibfnamefont{F.~T.} \bibnamefont{Smith}},
  \bibinfo{journal}{J. Fluid Mech.} \textbf{\bibinfo{volume}{113}},
  \bibinfo{pages}{407} (\bibinfo{year}{1981}).

\bibitem[{\citenamefont{Smith}(1985)}]{smith3}
\bibinfo{author}{\bibfnamefont{F.~T.} \bibnamefont{Smith}},
  \bibinfo{journal}{J. Fluid Mech.} \textbf{\bibinfo{volume}{171}},
  \bibinfo{pages}{263} (\bibinfo{year}{1985}).

\bibitem[{\citenamefont{Fornberg}(1985)}]{fornberg2}
\bibinfo{author}{\bibfnamefont{B.}~\bibnamefont{Fornberg}},
  \bibinfo{journal}{J. Comp. Phys.} \textbf{\bibinfo{volume}{61}},
  \bibinfo{pages}{297} (\bibinfo{year}{1985}).

\bibitem[{\citenamefont{Taneda}(1956{\natexlab{b}})}]{taneda2}
\bibinfo{author}{\bibfnamefont{S.}~\bibnamefont{Taneda}},
  \bibinfo{journal}{Journal of Physical Society of Japan}
  \textbf{\bibinfo{volume}{11}}, \bibinfo{pages}{1104}
  (\bibinfo{year}{1956}{\natexlab{b}}).

\bibitem[{\citenamefont{Johnson and Patel}(1999)}]{johnson}
\bibinfo{author}{\bibfnamefont{T.~A.} \bibnamefont{Johnson}} \bibnamefont{and}
  \bibinfo{author}{\bibfnamefont{V.~C.} \bibnamefont{Patel}},
  \bibinfo{journal}{J. Fluid Mech.} \textbf{\bibinfo{volume}{378}},
  \bibinfo{pages}{19} (\bibinfo{year}{1999}).

\bibitem[{\citenamefont{Koumoutsakos and Leonard}(1995)}]{leonard}
\bibinfo{author}{\bibfnamefont{P.}~\bibnamefont{Koumoutsakos}}
  \bibnamefont{and} \bibinfo{author}{\bibfnamefont{A.}~\bibnamefont{Leonard}},
  \bibinfo{journal}{J. Fluid Mech.} \textbf{\bibinfo{volume}{296}},
  \bibinfo{pages}{1} (\bibinfo{year}{1995}).

\bibitem[{\citenamefont{Kawamura et~al.}(1985)\citenamefont{Kawamura, Takami,
  and Kuwahara}}]{kuwahara}
\bibinfo{author}{\bibfnamefont{T.}~\bibnamefont{Kawamura}},
  \bibinfo{author}{\bibfnamefont{H.}~\bibnamefont{Takami}}, \bibnamefont{and}
  \bibinfo{author}{\bibfnamefont{K.}~\bibnamefont{Kuwahara}},
  \bibinfo{journal}{Fluid Dynamic Res.} \textbf{\bibinfo{volume}{1}},
  \bibinfo{pages}{145} (\bibinfo{year}{1985}).

\bibitem[{\citenamefont{Saad}(2004)}]{saad}
\bibinfo{author}{\bibfnamefont{Y.}~\bibnamefont{Saad}},
  \emph{\bibinfo{title}{Iterative Methods for Sparse Linear Systems}}
  (\bibinfo{publisher}{SIAM}, \bibinfo{year}{2004}), \bibinfo{edition}{2nd} ed.

\end{thebibliography}
\end{document}